\documentclass[a4paper]{article}
\usepackage[utf8]{inputenc}

\usepackage{hyperref}
\usepackage{booktabs}
\usepackage{subfigure, dsfont, multirow, array}
\usepackage{graphicx}
\usepackage[a4paper, margin=0.8in]{geometry}
\usepackage[11pt]{moresize}

\title{A Systematic Analysis of Fine-Grained Human \\Mobility Prediction with On-Device Contextual Data}
\author{
    Huoran Li, {\small Peking University, lihuoran@pku.edu.cn}
}

\begin{document}

\maketitle
\date{}

\begin{abstract}
User mobility prediction is widely considered to be helpful for various sorts of location based services on mobile devices. A large amount of studies have explored different algorithms to predict where a user will visit in the future based on their current and historical contexts and trajectories. Most of them focus on specific targets of predictions, such as the next venue a user checks in or the destination of her next trip, which usually depend on what their task is and what is available in their data. While successful stories are often reported, little discussion can be found on what happens if the prediction targets vary: whether coarser locations are easier to be predicted than finer locations, and whether predicting the immediate next location on the trajectory is easier than predicting the destination. On the other hand, commonly used in these prediction tasks, few have utilized finer grained, on-device user behavioral data, which are supposed to be indicative of user intentions. In this paper, we conduct a systematic study on the problem of mobility prediction using a fine-grained real-world dataset. Based on a Markov model, a recurrent neural network, and a multi-modal learning method, we perform a series of experiments to investigate the predictability of different types of granularities of prediction targets and the effectiveness of different types of signals. The results provide many insights on what can be predicted and how, which sheds light on real-world mobility prediction in general.
\end{abstract}


\section{Introduction}
\label{sec:introduction}

Human mobility prediction has drawn increasing attentions in the past a few years. 
The ability to predict the next location of a user is widely believed to be 
beneficial for many applications and services, including but not limited to smart transportation, personalized service recommendation, public resource management, and so on. Up to now, a large amount of mobility prediction methods have been proposed, ranging from pattern-based methods~\cite{ying2013mining, ozer2016predicting, monreale2009wherenext, ying2011semantic, cao2017efficient}, to Markov model-based methods~\cite{chen2014nlpmm, gambs2012next, ye2013s, asahara2011pedestrian, yu2015location, liu2016novel, li2015hybrid, huang2017mining, jiang2016predicting, zhang2016gmove, mathew2012predicting, chen2016mining}, and to deep neural networks~\cite{liu2016predicting, feng2018deepmove, yao2017serm, wu2017spatial}. 
These models are applied to various scenarios, including indoor walking~\cite{asahara2011pedestrian}, venue recommendation~\cite{zhang2016gmove}, urban commuting~\cite{feng2018deepmove}, or even intercontinental trips~\cite{liu2016predicting}. Successful stories are often reported, with improved accuracy numbers on particular prediction targets. 

Despite these continuously advanced models and improved results, 
some fundamental questions of mobility prediction have never been answered, or even discussed systematically. 

First, does the granularity of prediction targets matter?   
In practice, the granularity of the next location that a system can predict is critical for the feasibility of real world applications. 
For example, if we want to recommend a point of interest to a tourist, we only need to know the next city they are going to, e.g., Central Park is a reasonable recommendation for most people coming to the New York city soon. However, if the goal is to recommend a restaurant, knowing only the next city is not enough, and a prediction at a finer granularity would be needed (e.g., recommending a nearby restaurant if the user is predicted to come to Central Park). 
Given the same signals (e.g., behavioral data on their smartphones), how much harder is it to predict Central Park versus Manhattan? And does it require a more sophisticated model? Existing studies usually concentrate on a particular type of targets, mostly due to their task or the data they have access to (e.g., check-in logs), and few have taken into account the impact of the granularity of their prediction targets (locations).

Second, does the salience (or meaningfulness) of the next location matter? Almost all existing studies are aimed to predict 
the \textbf{exact} next location that a user is going to access (at a fixed granularity), and they do not distinguish the intention of the visit. For example, they care about whether ``\textit{the next location on the user's trajectory is a coffee shop}'', 
but usually do not ask whether the coffee shop is a temporary stop or the destination of her trip. 
In practice, user's movements are continuous, and predicting the next ``meaningful'' location is much more useful than predicting the ``very next'' location. But how to define the meaningfulness of a location? It may be figured out when the user's activities at each location are available; when they are not, the salience of a location can be inferred from how long the user stays there. 
Is predicting the next sustainable location harder than predicting the next any location? Does it require special solutions?   


Last but not the least, how much do different behavioral signals matter?  Existing studies usually build the prediction model based on features related to (the fixed granularity of) locations, such as historical locations, their timestamps, and semantic tags of locations. Indeed, these features are naturally predictive of 
future locations at the same granularity. However, 
when comes to predict next locations at a different granularity, or locations with different intents/duration of stay, are these features still effective?  On the one hand, in addition to 
these ``location records,'' there may be many other signals from the user's behaviors, which may be useful, or even more indicative of the user's intent or trajectory.  
For example, different types of usage behaviors and system status can be collected from the user's mobile devices, at different granularities, which can be predictive of the future actions of the user including their next locations. 
These fine-grained behavioral signals are usually neglected in existing studies, mostly because they were not available. Are these behavioral signals useful at all for mobility prediction? Do they add value to trajectory data? Does their effectiveness vary for different prediction targets?

We take the initiative to bridge this gap by addressing these questions. We conduct a systematic analysis of the task of predicting the next location of mobile users. Instead of trying to find the best model for a particular setup (as in most existing studies), we focus on comparing and analyzing different setups: different \textbf{granularities} and \textbf{salience} (duration of stay) of the target locations, differerent \textbf{prediction models}, and different types of \textbf{behavioral signals} as features. 
This comprehensive study is enabled by a recently collected large-scale dataset of real-world usage of mobile devices. 

The main goal of this paper is to understand how the above factors influence the performance of mobility prediction. 
We do not aim to find the best model for particular applications, as most existing studies do, and instead the results of our analysis provide insights on the feasibility and how to optimize mobility prediction for particular applications.  
The major contributions of our work are as follows:

\begin{itemize}
    \item We propose the first systematic study on how the variations of problem setup can influence the performance of human mobility prediction. To be more concrete, our paper discusses the impact of the granularity and duration of stay of target locations, as well as different behavioral features on the prediction accuracy.
    \item We carefully design an empirical experiment to analyze the impact of the above three factors. Based on a comprehensive, multi-grained,  real-world dataset, we conduct a series of measurements, qualitatively and quantitatively, to address the above questions. The results demonstrate many interesting patterns of user mobility and reveal many useful insights.
    \item We provide design implications derived from our measurements, which can guide the building of applications of mobility prediction in practice.
\end{itemize}

The rest parts of this paper is organized as follows. We first introduce papers related to user mobility prediction, and discuss the current research status on the above three issues in Section~\ref{sec:related}. Then, we introduce the scope of this paper and our analysis pipeline in Section~\ref{sec:problem}. After that, we introduce the dataset that is used for the measurement study in Section~\ref{sec:dataset}. Details of the experiments and results of the three steps of our study are presented in Section~\ref{sec:granularity}, Section~\ref{sec:target}, and Section~\ref{sec:feature}, respectively. After giving a several implications in Section~\ref{sec:implication}, we finally conclude our work in Section~\ref{sec:conclusion}.
\section{Related Work}
\label{sec:related}

User mobility prediction is increasingly finding its place in the past few years. Researchers have already proposed a variety of prediction models based on multiple technologies, including pattern-based models~\cite{ying2013mining, ozer2016predicting, monreale2009wherenext}, Markov-based models~\cite{zhang2016gmove, chen2014nlpmm, gambs2012next}, and neural network models~\cite{liu2016predicting, feng2018deepmove, yao2017serm}. However, the goal of most existing studies is just to optimize the model under a fixed problem setup, i.e., a fixed location granularity, a particular target salience, and a specific set of features. We categorize the related literature from the preceding aspects.

\subsection{Location Granularity}

In most cases, the location granularity is determined by the prediction task or the data that existing studies can access. Generally, location data have three forms. A ``location'' is actually a point of interest (POI) (e.g., check-in data~\cite{liu2016predicting, cao2017efficient, zhang2016gmove, ye2013s, li2015hybrid, huang2017mining, feng2018deepmove, yao2017serm, cheng2013you, jia2016location}), a connected region (e.g., a region covered by a base station~\cite{yu2015location, liu2016novel, ozer2016predicting} or a surveillance camera~\cite{chen2016mining, chen2014nlpmm}, or a pair of coordinates (e.g., GPS coordinates~\cite{mathew2012predicting, monreale2009wherenext, ying2011semantic, gambs2012next, asahara2011pedestrian, ying2013mining, wu2017spatial}). For POI data, the location granularity is fully determined by the granularity of POIs. For regional data, the location granularity refers to the average size of all regions. For real-value coordinates, existing studies usually convert continuous data into discrete regions. Therefore, the granularity of locations essentially refers to the size of regions. Once the data processing is done, the granularity will never change again.

As stated previously, the granularity of the location to be predicted is critical for the feasibility of real-world applications. However, existing studies usually design their models based on a fixed granularity, and seldom make the in-depth analysis on how location granularity can influence the prediction performance. Although there are some studies that involved more than one datasets with  different location granularities, they still did not carefully compare the impact of granularity on the model's performance. For example, Liu et al.~\cite{liu2016predicting} used both Gowalla Dataset~\cite{cho2011friendship} and Global Terrorism Database~\cite{bhargava2015and}, but just observed the results on these two datasets, respectively. The authors did not make further research on how will the location granularity affect the prediction performance.

\subsection{Target Salience}

In practice, not every location is meaningful and worth predicting as the prediction target. Since user's trajectory traces are continuous, before reaching the real destination, the user can pass by many medium points. Predicting the user's real destination in the future is more practically meaningful, rather than those who she passes by only. Hence, we should carefully take into account identifying each location's salience. Only a location whose salience is long enough shall be considered as the meaningful prediction target. 

Unfortunately, to the best of our knowledge, almost all existing studies did not make in-depth considerations of this issue. They did not consider the concrete salience of each location, and hence did not distinguish which locations are worthy predicting according to location salience. Consequently, existing studies usually just took the ``exact'' next location that the user will visit as the prediction target. Although some efforts such as Ozer et al.~\cite{ozer2016predicting}  defined three different prediction targets and compared the performance under different targets, the impact of salience is still primitive and not comprehensive enough. Therefore, we claim that it is quite important to make more detailed exploration on this issue.

\subsection{Involved Features}

Intuitively, features that are most relevant to mobility prediction are users' historical locations and the corresponding timestamps. Almost all existing studies involve these two kinds of information into their prediction model. In addition, in order to better understand the semantic meanings of locations, some studies also involved semantic tags of locations in the model~\cite{ying2011semantic, zhang2016gmove, ying2013mining, yao2017serm, wu2017spatial, cheng2013you}. Indeed, these three kinds of features are all closely related to user's movement, so they are naturally predictive for future locations. However, designing a mobility prediction model based on only these three kinds of features is far from sufficient and satisfactory. In addition to these ``location records'', there can be many other signals that possibly indicate user's interests and behaviors on a specific location. These signals, including usage logs and system status, are rather  useful, or even more indicative of the user's intent or movement. Unfortunately, due to the lack of such type of data, existing studies seldom tried these fine-grained behavioral features. We are interested about whether these features do bring unique value, and if yes, how about their effectiveness vary against different prediction targets.

\section{Problem Statement And Analysis Pipeline}
\label{sec:problem}

As stated in the introduction, the theme of this paper is exploring the impact of problem setup on prediction accuracy in mobility prediction scenarios. To be more specific, a problem setup consists three components: location granularity, target salience, and input features. To clarify the scope of our work, in this section, we first present a formal description of the related concepts and the mobility prediction task, and then introduce the research questions we want to answer.

\subsection{Mobility Prediction Formulation}

\noindent \textbf{DEFINITION 1} (Location). A location $l$ is defined as a region of connected area. Each location is identified by a numerical ID.

\smallskip

\noindent \textbf{DEFINITION 2} (Location Granularity). Location granularity $G$ is the average area of all locations. We say that the locations are fine-grained with a small average area, or coarse-grained with a large average area. $l^G$ denotes a location at the location granularity $G$. We ignore the superscript when it does not cause ambiguity.

\smallskip

\noindent \textbf{DEFINITION 3} (Location Record). A location record $r$ is a tuple of a timestamp $t$ and a location identification $l$, i.e., $r^G=(t, l^G)$. A location record could tell us where the user is at a specific moment. We also ignore the superscript here when it does not cause ambiguity.

\smallskip

\noindent \textbf{DEFINITION 4} (Trajectory). Given a user $u$ and a time window $w$, a trajectory is a sequence of location records $T_w^u=r_i r_{i+1} ... r_{i+k}$, which illustrates the user's movement in a period of time.

\smallskip

\noindent \textbf{DEFINITION 5} (Staying Time \& Location Salience). A location's staying time is defined as the duration that the user stays in this location. Formally, If the user enters a location $l_j$ at time $t_1$ and leaves at time $t_2$, then the staying time of $l_j$ is $S_{l_j}=t_2-t_1$. The location salience refers to the importance of a location. In this paper, we define that the location salience is positively related with the staying time. The longer the stay, the higher the salience. From this point of view, we could use a location's staying time to represent its salience.

\smallskip

\noindent \textbf{DEFINITION 6} (Target Location). The target location $l_t$ is defined as the very  \textbf{first} location whose salience is high enough in the user's future trajectory. Whether a salience is high enough depends on a specific criterion $C$.

\smallskip

\noindent \textbf{DEFINITION 7} (Mobility Prediction). The goal of mobility prediction is predicting $l_t$ based on the user's historical trajectory and usage behaviors. Formally, given a historical time window $w_h$, a future time window $w_f$, and a feature set $F$ that corresponds to $T_{w_h}^u$, the goal is to  predict $l_t$ that is selected from $T_{w_f}^u$ (according to $C$) based on $F$. 

\subsection{Research Questions}

With the theme of user mobility prediction, the research questions that we want to answer include three aspects:

\smallskip

\textbf{\textit{RQ1: What is the predictability of the mobility prediction task under different location granularities?}} In other words, we would like to know how the location granularity $G$ can affect the prediction accuracy.

\smallskip

\textbf{\textit{RQ2: What is the impact of target salience on prediction performance?}} It refers that we would like to investigate the trends of the prediction accuracy with the varying  selection criterion $C$.

\smallskip

\textbf{\textit{RQ3: What is the significance of multiple usage features?}} On the one hand, we want to know the predictive power of multiple types of usage features. On the other hand, we also want to identify whether and how the features' effectiveness could vary for different prediction targets.

\subsection{Analysis Pipeline}
\label{sec:pipeline}

To systematically explore the research questions introduced in the previous section, we propose an analysis pipeline that could systematically measure the impact of problem setups on prediction performance. 
The pipeline consists of four steps: 1) Data Pre-processing, 2) Location Granularity Analysis, 3) Target Salience Analysis, and 4) Behavioral Feature Analysis. The details of every single step are presented as follows.

\subsubsection{Data Pre-processing.} As the first step of our pipeline, we extract user trajectories and the related usage data from the raw dataset. The extraction can be done in multiple ways. For example, a possible method is considering location records within a single day as a trajectory. After the trajectories are extracted, we then collect usage data that associated to each trajectory, such as user behaviors, smartphone's status, context, and so on.

\subsubsection{Location Granularity Analysis.} Once the trajectories and the associated data are ready, we then conduct a descriptive analysis to present a intuitive understanding of the data. Through this step, we demonstrate several basic statistics of the data, such as transition probabilities between locations, average staying time in each location, and so on. Moreover, to quantitatively verify the feasibility of prediction under each location granularity, we build a group of machine learning models based on the historical part of the trajectories, and then use the models to predict a user's future movement to see how will the models perform under each location granularity.

\subsubsection{Target Salience Analysis.} In this step, we define several criteria to select the target location, and then investigate the impact of salience on prediction accuracy. In this step, we also adopt state-of-the-art machine learning models to make the predictions, and then compare the prediction performance of the models under different prediction criteria.

\subsubsection{Behavioral Feature Analysis.} In the last step of our pipeline, we explore the predictive power of multiple types of behavioral features. To be more specific, we will try various kinds of feature combinations, to see which of them can significantly contribute to the mobility prediction task. Since different kinds of features have different forms, in this step we need to adopt a multi-modal learning method to synthesize different features. 

\section{The Dataset}
\label{sec:dataset}

\begin{figure}
	\centering
	\begin{center}
		\includegraphics[width=0.45\textwidth]{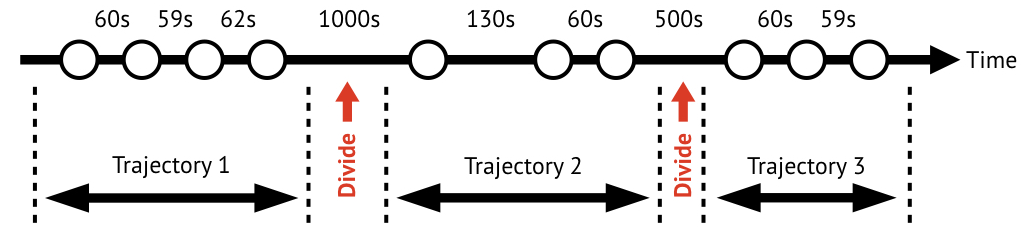}
		\caption[7.5pt]{An example of trajectory extraction. The time intervals pointed by red arrows are larger than five minutes (300 seconds), so the entire sequence is divided into three trajectories from these two places.}
		\label{fig:trajectory-extraction}
	\end{center}
\end{figure}

We have the access to the Sherlock dataset~\cite{mirsky2016sherlock}, which is a long-term and comprehensive data collections maintained by the BGU Cyber Security Research Center. In this section, we briefly introduce this dataset, and explain how we extract related data from the raw dataset. More details of this dataset can be found on the official website~\footnote{More details of Sherlock dataset can be accessed via \url{http://bigdata.ise.bgu.ac.il/sherlock/}}.

\subsection{Dataset Overview}

Essentially, the Sherlock dataset is a multidimensional time-series recording nearly \textbf{all software signals and hardware sensors} that can be obtained from a Samsung Galaxy S5 smartphone without root privileges. The creators of the dataset recruited a group of volunteers, then provided each of the volunteers with a Samsung Galaxy S5 smartphone and asked the volunteers to use this phone as their major device. Each phone was pre-installed a data collection agent (an Android app). During the volunteers usage, their usage data were recorded by the agent with their consensus. In this paper, we adopt data spanning three months (July 1st, 2016 to September 30th, 2016) and covering 51 participants.

The agent collects data in two ways: active collection and passive collection. As the name suggests, active collection means that the agent reads information and records it periodically, while passive collection means that the agent will make a record when an event occurs (for example, when a phone call comes in).

\subsection{Location Data}

To protect privacy of the volunteers, in this dataset, the exact geo-locations of the volunteers are anonymized. Instead, the creators of the dataset performed a K-Means clustering algorithm over all the volunteers' occurrences. Only the cluster IDs of the users' locations are reported. These IDs do not contain any geographic information. They can be used only as categorical identifiers. Through this way, the entire user movement range is divided into several locations. 

Due to the data anonymization, we do not really know the actual size of locations at each granularity. In other words, the location granularity in this dataset is only a relative concept. The more clusters, the finer the granularity. Since the location data under different granularities come from the same raw geo-locations, we can fairly compare the prediction performance under different data granularity without interfered by other factors. This makes the Sherlock dataset quite suitable for the location granularity analysis in our pipeline. 

In the Sherlock dataset, there are six different settings of $M$ (number of clusters) in the K-Means clustering: 5, 10, 25, 50, 75, and 100. The location records are actively collected at a frequency of around once per minute. Each location record contains six IDs that corresponds to the location ID under six different $M$s, respectively.

\subsection{Usage Data}
\label{sec:usage-data}

The Sherlock dataset contains rich usage information of users. However, not all of them are suitable for the mobility prediction. Some features are not practically usable because they are too sparse, such as SMS log, call log, app changing log, and so on. Other features are unusable because intuitively they are not relevant to user's movement, such as screen brightness and speaker volume. In this study, we select three groups of usage features: 

\begin{itemize}
    \item \textit{App usage data}. App usage data are actively collected. We could know what apps are running (both in foreground and background) for every five seconds. 
    \item \textit{Location sensor data}. This parts of data include many sensors that are related to smartphone's motion and gesture, such as accelerometer. They are actively recorded for every 15 seconds. Totally, there are 238 readings.
    \item \textit{Broadcast data}. Whenever an Android system broadcast is sent, its content will be passively recorded by the agent. In the Sherlock dataset, there are 82 kinds of broadcasts.
\end{itemize}

App usage data can demonstrate the users' active usage behavior because app usage is the most important form of usage. The other two groups of features could represent the device's system status. In particular, they can also cover geographically related system status (e.g., speed and acceleration of the smartphone). Therefore, we choose the above three groups of usage features, and we believe these features can form an informative and representative description of users' usage from multiple aspects.

\subsection{Trajectory Extraction}


As the last part of the dataset introduction, we examine whether we can construct usable trajectories from the location records. Ideally, we can arbitrarily intercept any part from a user's location sequence as a trajectory. However, it is not a good strategy in practice because there exist missing records in the raw dataset. Such a case happens either when the device was powered off or when the agent failed to record data. In either cases, the time interval between two consecutive records might be much longer than one minute. When this happens, we do not have enough information about the user movement during this long period of time, so we discard data within this time period by dividing the location sequence into two trajectories from this point. In this study, we set this threshold as five minutes, i.e., each pair of consecutive location records that are closer than five minutes will be put in the same trajectory. Figure~\ref{fig:trajectory-extraction} presents an example of the trajectory extraction. Since the sampling interval is about one minute, the number of records in a trajectory is approximately equal to the duration of the trajectory (in minute).

To make meaningful user mobility predictions, we filter out trajectories that are too short. In this paper, this threshold is set to be one hour, as we think that trajectories shorter than a hour can not perform sufficient information to understand the user's movement. Finally, we obtained \textbf{4,785} trajectories. The extracted trajectories keep the original form of the raw data (i.e., actively sampled records), so there might be consecutive duplicated locations in a trajectory. 

\section{Location Granularity Analysis}
\label{sec:granularity}

So far, we have proposed the research questions, the analysis pipeline, and the real-world dataset. Now we can conduct our  study over this dataset according to the pipeline. As the first part of the analysis, in this section, we discuss the impact of location granularity on the mobility prediction task. We begin with a descriptive analysis on the location granularity, to see some basic characteristics of the data, and then make a quantitative analysis to rigorously compare the prediction performance under different location granularities.

\subsection{The Descriptive Analysis}

As stated previously, the location granularity in Sherlock dataset is a relative concept, i.e., we do not know the actual scale of locations at each granularity. In order to have a more intuitive understanding of each granularity, we begin with exploring the location data through a descriptive analysis. We first calculate the average staying time in a location under different location granularities (listed in Table~\ref{tab:staying-time}). First of all, we can see that the average staying time is rapidly decreasing with the increment of $M$, as expected. An average staying time that is less than three minutes does not seem to meet our common sense, because usually users will not move that frequently. On the other hand, an average staying time that is longer than an hour is also not consistent with our common knowledge. This means that a user will visit less than 24 locations in a single day, which is likely to be less than our habits. Therefore, a ``location'' under $M=25$ or $M=50$ is more likely to be consistent with our common sense.

\begin{figure}
    \centering
    \begin{center}
        \subfigure[$M=25$\label{fig:transition-025}]
        {\includegraphics[width=0.3\textwidth]{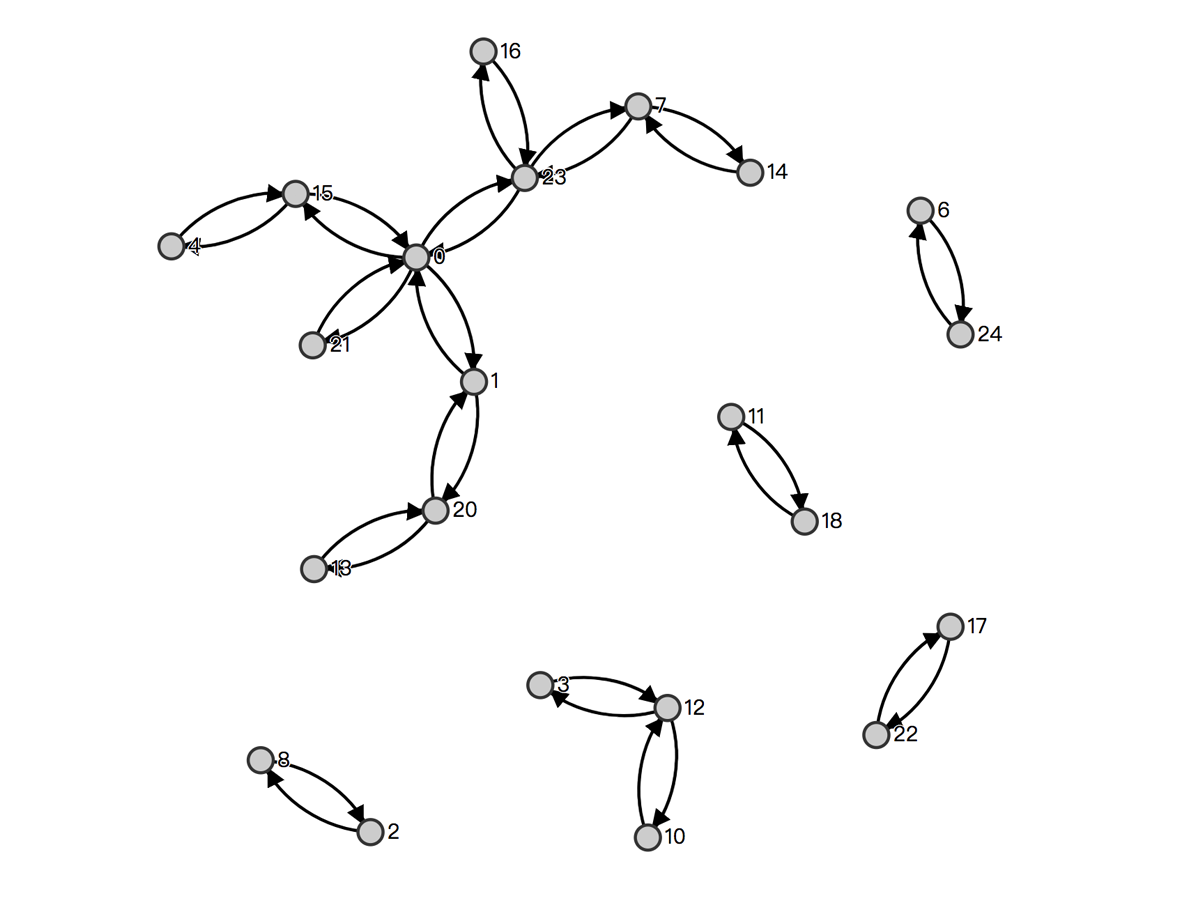}}
        \subfigure[$M=50$\label{fig:transition-050}]
        {\includegraphics[width=0.3\textwidth]{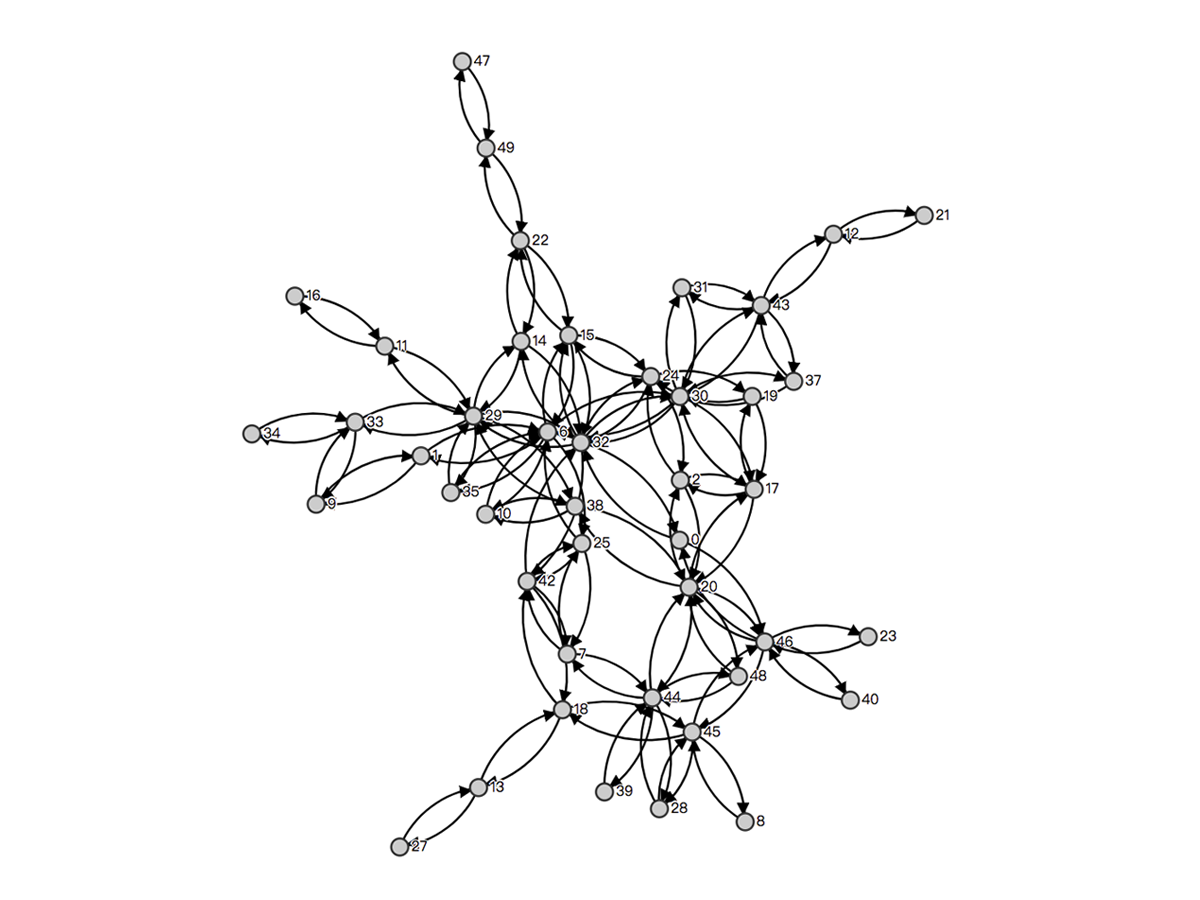}}
        \subfigure[$M=75$\label{fig:transition-075}]
        {\includegraphics[width=0.3\textwidth]{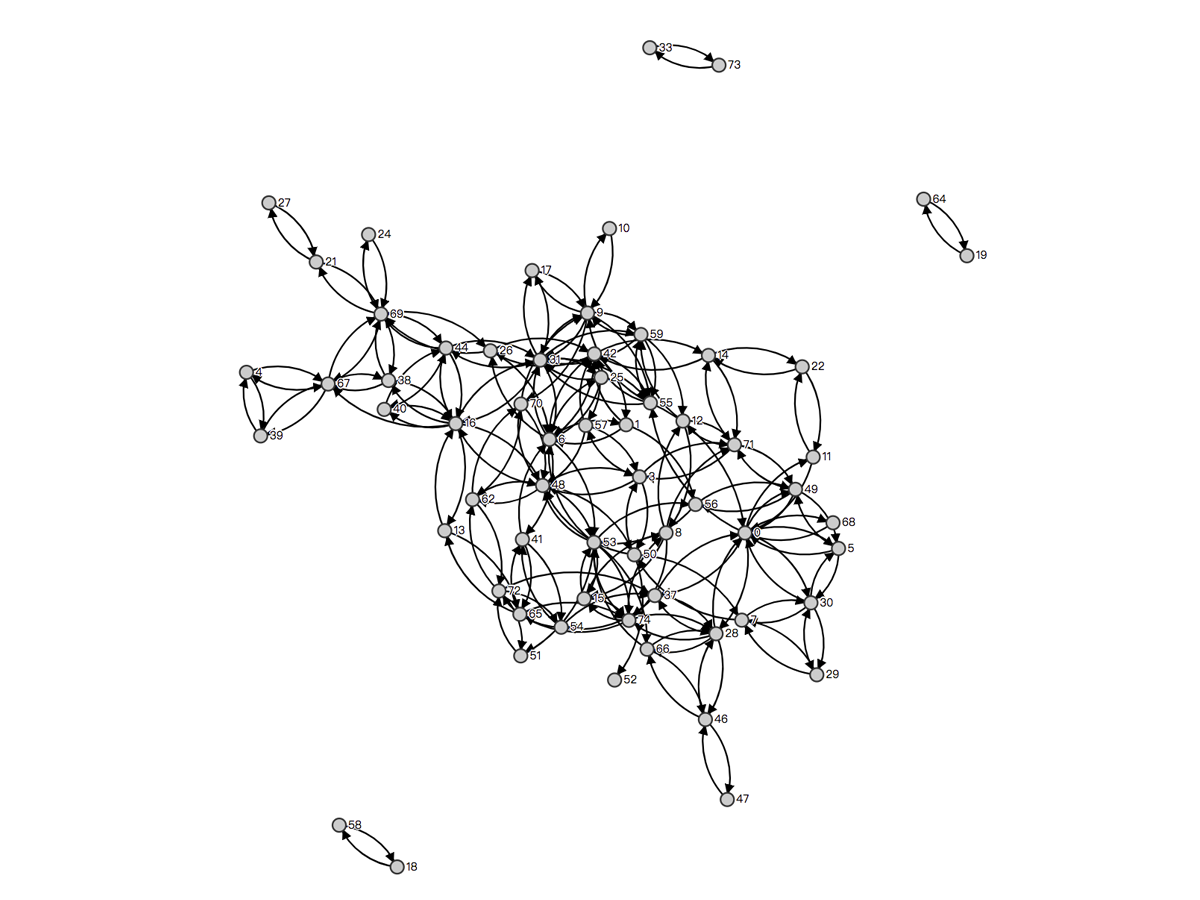}}
        \caption[7.5pt]{The transition diagrams under three location granularities.}\label{fig:transition}
    \end{center}
\end{figure}

We then explore the transitions between different locations under different location granularities. For each pair of locations $<l_i, l_j>$, we count the number of transitions from $l_i$ to $l_j$, then construct a transition matrix with the shape $M*M$. After that, we discard transitions that occurred too few and use the remained transitions to draw a directed acyclic graph (DAG). Due to page limit, we show the DAGs under $M=25$, $50$, and $75$ in Figure~\ref{fig:transition}. When $M=25$, the diagram is made up of various separated clusters, indicating that users tend to move within an small region. When $M=50$, all the locations are connected into a big group, but the structure of the transitions are still clear. When $M=75$, however, the transition diagram becomes too dense and a little bit messy, so it is less likely to happen in reality. As a summary, from the transition aspect, we verified the previous conclusion again: $M=25$ or $M=50$ are more likely to be reasonable definitions from this aspect.

\begin{table}[htbp]
\centering
\caption{The average staying time ($Avg.Stay$, in minutes) in a location under different location granularities.}
\label{tab:staying-time}
\begin{tabular}{c|cccccc}
\hline
\textbf{$M$} & 5 & 10 & 25 & 50 & 75 & 100 \\
\hline
\textbf{$Avg.Stay$} & 161.22 & 75.62 & 10.02 & 3.99 & 2.76 & 2.42 \\
\hline
\end{tabular}
\end{table}

\subsection{Quantitative Analysis}

The above descriptive analyses provide some intuitive impressions about each location granularity. In the second part of this section, we conduct a quantitative experiment to rigorously compare the predictability under different location granularities. In a nutshell, in this experiment we simulate a group of mobility prediction queries based on the user trajectories, then train a prediction model, and finally analyze the prediction accuracy under each location granularity. We begin with the query simulation.

\subsubsection{Query Simulation}

\begin{table}[htbp]
\centering
\caption{Sizes of the testing set under each location granularity and each target selection criterion.}
\label{tab:testing-size}
\begin{tabular}{l|rrrrrr}
\hline
\textbf{Target Selection} & \multicolumn{6}{c}{\textbf{\# of Locations ($M$)}} \\
\cline{2-7}
\textbf{Criterion} & 5 & 10 & 25 & 50 & 75 & 100 \\
\hline
Successive   & 849 & 1,429 & 2,425 & 2,968 & 3,078 & 3,182 \\
Important@2  & 824 & 1,368 & 2,242 & 2,813 & 2,945 & 3,050 \\
Important@5  & 796 & 1,276 & 1,973 & 2,406 & 2,476 & 2,551 \\
Important@10 & 739 & 1,180 & 1,711 & 1,917 & 1,864 & 1,919 \\
Longest@3    & 208 & 740   & 1,935 & 2,640 & 2,811 & 2,933 \\
Longest@5    & 93  & 354   & 1,596 & 2,392 & 2,631 & 2,786 \\
Longest@10   & 37  & 146   & 1,132 & 1,993 & 2,284 & 2,447 \\
\hline
\end{tabular}
\end{table}

To simulate a location prediction query, we randomly select a time point from the trajectory, and then divide the trajectory into two parts by this time point: the first part is considered as the historical trajectory of the user ($T_{w_h}^u$), and the second part is the future trajectory of the user ($T_{w_f}^u$) that contains the target location $l_t$ we want to predict. Such a pair $<T_{w_h}^u, T_{w_f}^u>$ can represent a query that happens at the division moment. In practice, suppose there are $n$ location records in a trajectory. We randomly choose a positive integer $m$ ($m < n$), then consider the first $m$ records as $T_{w_h}^u$ and the rest of the records as $T_{w_f}^u$. The location prediction request is occurred between $T_{w_h}^u$ and $T_{w_f}^u$, so the last location in $T_{w_h}^u$ can be seen as the current location of the user. In order to avoid creating a pair that is too unbalanced, we restrict that both $T_{w_h}^u$ and $T_{w_f}^u$ contain at least 20\% of the original trajectory. Moreover, to augment our data, we make five individual simulations on one trajectory. Therefore, we get 23,925 simulated queries from the 4,785 trajectories.

The experiment here follows the standard machine learning pipeline, so we split all the 23,925 queries into three groups: a training set, a validation set, and a testing set. To keep the independence of these three sets, we preserve that the five queries generated from the same trajectories will be put into the same set. Then, for each simulated query, we extract the next successive location of the user from the query's $T_{w_f}^u$ as the target location $l_t$. The next successive location of the user is the first location in $T_{w_f}^u$ that is not equal to the current location. For example, if the current location is $A$, and the location list in $T_{w_f}^u$ is $(A,A,B,B,C,A)$, then the next successive location of this query is $B$. However, according to this definition, there might not be a next successive location in some queries, i.e., the user never changes her location within $w_f$. In this case, these queries do not have a label to be predicted, so we discard these queries during this experiment. Therefore, we have less than 23,925 usable queries in this experiment, and the amount of usable queries varies under different location granularities. The sizes of the testing sets under different location granularities are listed in the first row of Table~\ref{tab:testing-size}.

\subsubsection{The Markov Model}

The first model we use here is a simple first-order Markov Model. We train the model based on all the queries' historical trajectories, and then use it to decide which location has the highest possibility to be visited based only on the current location. Notice that the next successive location cannot be the current location, so we do not count transitions that from a location to itself. In order words, the diagonal of the transition matrix of this Markov model contains only zero.

\subsubsection{The LSTM Model}

\begin{figure}
    \centering
    \begin{center}
        \subfigure[Trajectory pre-processing procedure.\label{fig:trajectory-preprocess}]
        {\includegraphics[width=0.4\textwidth]{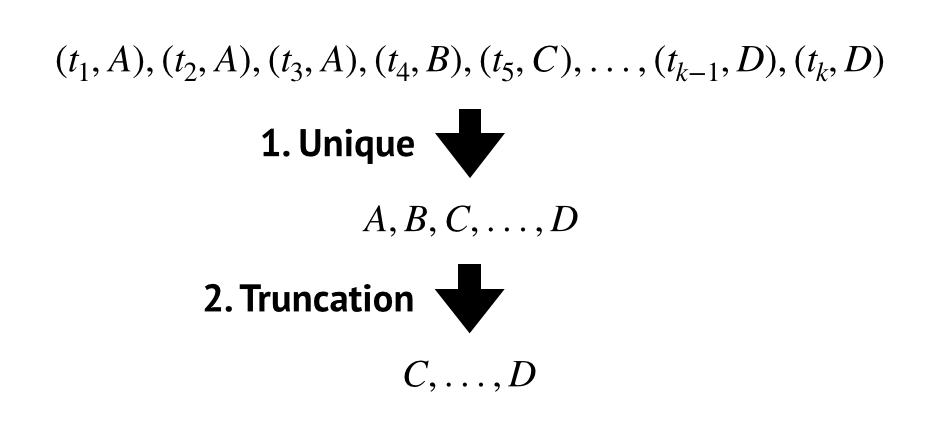}}
        \subfigure[The LSTM model.\label{fig:lstm-structure}]
        {\includegraphics[width=0.4\textwidth]{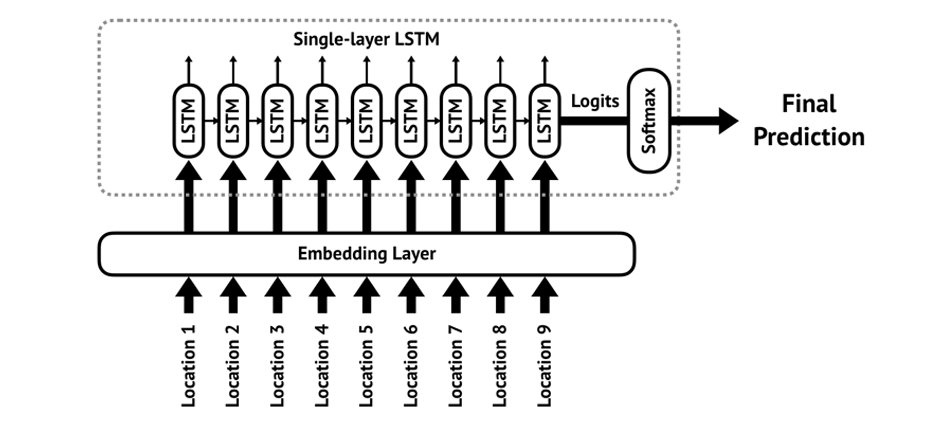}}
        \caption[7.5pt]{The trajectory pre-processing procedure and the structure of the LSTM model.}\label{fig:lstm}
    \end{center}
\end{figure}

Building a first-order Markov model to make predictions is rather easy and straightforward. However, it is too naive to capture the complex sequential information from the historical trajectories. As discussed in Section~\ref{sec:related}, there are already many RNN-based models for mobility prediction. Following the same principle, here we also use a LSTM model to learn from the historical trajectories and predict the next successive location. Since the location ID is a categorical value, we are actually implementing a LSTM classifier. 

We first pre-process each trajectory to make it more suitable for the LSTM model. The two-stage procedure is demonstrated in Figure~\ref{fig:trajectory-preprocess}. For each trajectory, we merge repeated consecutive locations into one (the ``Unique'' step), indicating that we care about where the user has been but not the duration time of stay. Then, for the simplified trajectory, we use only the latest 100 locations (the ``Truncation'' step), because the information that is too early can help very little. The structure of the LSTM model is illustrated in Figure~\ref{fig:lstm-structure}. We take each location record as a time step, while each location is embedded by an embedding layer before being fed into the model. The output of the last layer is logit, a $M$-dimensional vector, and then a Softmax layer is applied to generate the final prediction. When training the model, we adopt the \textit{cross-entropy} as our loss function. After a large amount of attempts, we get the best performance under the above settings.

\subsubsection{Experiment Results}

\begin{figure}
	\centering
	\begin{center}
		\includegraphics[width=0.41\textwidth]{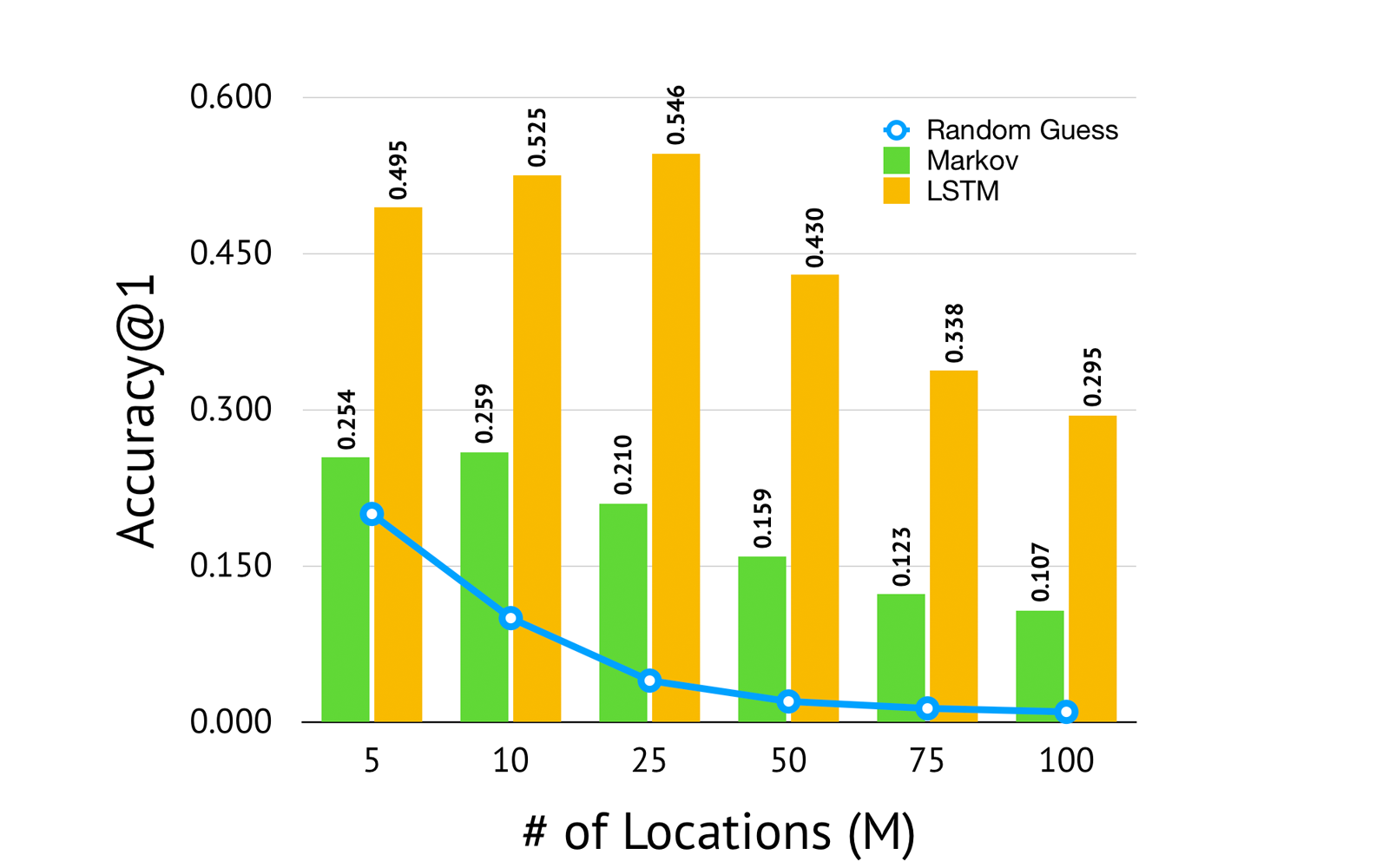}
		\caption[7.5pt]{Accuracy@1 of the next successive location under different location granularities. The green and yellow bars represent the Accuracy@1 of the Markov model and the LSTM model, respectively. We use the random guess as the baseline under each $M$, which is represented by the blue curve (i.e., Accuracy@1=$\frac{1}{M}$).}
		\label{fig:result-granularity}
	\end{center}
\end{figure}

The experiment results are shown in Figure~\ref{fig:result-granularity}. We use Accuracy@1 to evaluate the model performance. As for the Markov model (green bars), when $M=5$ or $M=10$, the results are quite close, which are 0.254 and 0.259, respectively. Besides, the prediction accuracy decreases with the growth of $M$, from 0.210 to 0.107. Intuitively, this result can be expected, because the user movement is more difficult to predict if the location is defined in a more fine-grained location. Through the performance of the random guess (the blue curve in Figure~\ref{fig:result-granularity}), we can also observe this phenomenon quite clearly. Therefore, for the Markov model, we can conclude that the prediction accuracy is lower with more fine-grained locations.

As for the LSTM model (yellow bars), however, we find a different result. On the one hand, the LSTM model significantly outperforms the Markov model for all location granularities. This conveys to us that the sequential information contained in the historical trajectories can be captured much better by the LSTM model. In addition, the relative performance difference between the two models becomes larger along with the growth of $M$, which means the LSTM model is adequate at handling fine-grained locations. On the other hand, the prediction accuracy is not monotonically decreasing with the growth of $M$. The accuracy increases when $M\leq25$, and then begins to decrease. The best performance is obtained under $M=25$.

\subsection{Summary}

Through the descriptive analysis, we can have a intuitive understanding of the scale of each location granularity. The granularity of locations is likely to be proper when $M=25$ and $M=50$. Subsequently, we use a quantitative experiment to show that location granularity does have a significant impact on the  prediction accuracy, and the prediction performance peaks when $M=25$, indicating that the prediction model outputs with the best results when the granularity is moderate. The prediction accuracy will result in a decline if the locations are too coarse-grained or too fine-grained. 
\section{Target Salience Analysis}
\label{sec:target}

As the second step of our analysis, in this section, we focus on investigating the impact of target salience on prediction performance. In this paper, we define three different criteria to select the target location based on location salience (i.e., staying time), and explore the prediction performance under these criteria. The detailed definitions are:

\begin{enumerate}
    \item \textit{Successive.} We have already introduced this criterion in the previous part of this paper. It takes the first location in $T_{w_f}^u$ that is not equal to the current location as the prediction target. This criterion is the most common one that is adopted by almost all the existing studies. 
    \item \textit{Important@$K$.} As mentioned above, we think that a location with sufficient staying time is more likely to be an important location rather than just a pass-by point. Following this principle, \textit{Important@$K$ is defined as the first location in $T_{w_f}^u$ that the user stays for at least $K$ minutes}. Apparently, The threshold $K$ determines the standard of becoming an important location. Similarly, we also require that this location cannot be equal to the user's current location.
    \item \textit{Longest@$K$.} Defining an important location by manually setting a threshold may not be always reasonable, because users usually have very different movement patterns. For example, Alice might have a relatively long stay in every location, while Bob tends to stay in every location short. Thus, a unified threshold is not suitable for all users. Motivated by this, we define Longest@$K$, which refers to \textit{the location with the longest staying time among the first $K$ locations of $T_{w_f}^u$}. We still require that this location cannot be equal to the current location of the user.
\end{enumerate}

In this paper, we choose $K=2, 5, 10$ for Important@$K$ and $K=3, 5, 10$ for Longest@$K$. Similar to the Successive, the number of labeled queries under different target selection criterion are varied. The testing sizes of all these combinations are can be found in Table~\ref{tab:testing-size}. For Important@$K$, if there is not a location that the user stays for at least $K$ minutes in $T_{w_f}^u$, then there would not be a label for this query. For Longest@$K$, if the total number of locations in $T_{w_f}^u$ is less than $K$, then we do not have enough candidates, hence we cannot select a valid prediction target.

\subsection{Experiment Results}

\begin{figure}
	\centering
	\begin{center}
		\includegraphics[width=0.41\textwidth]{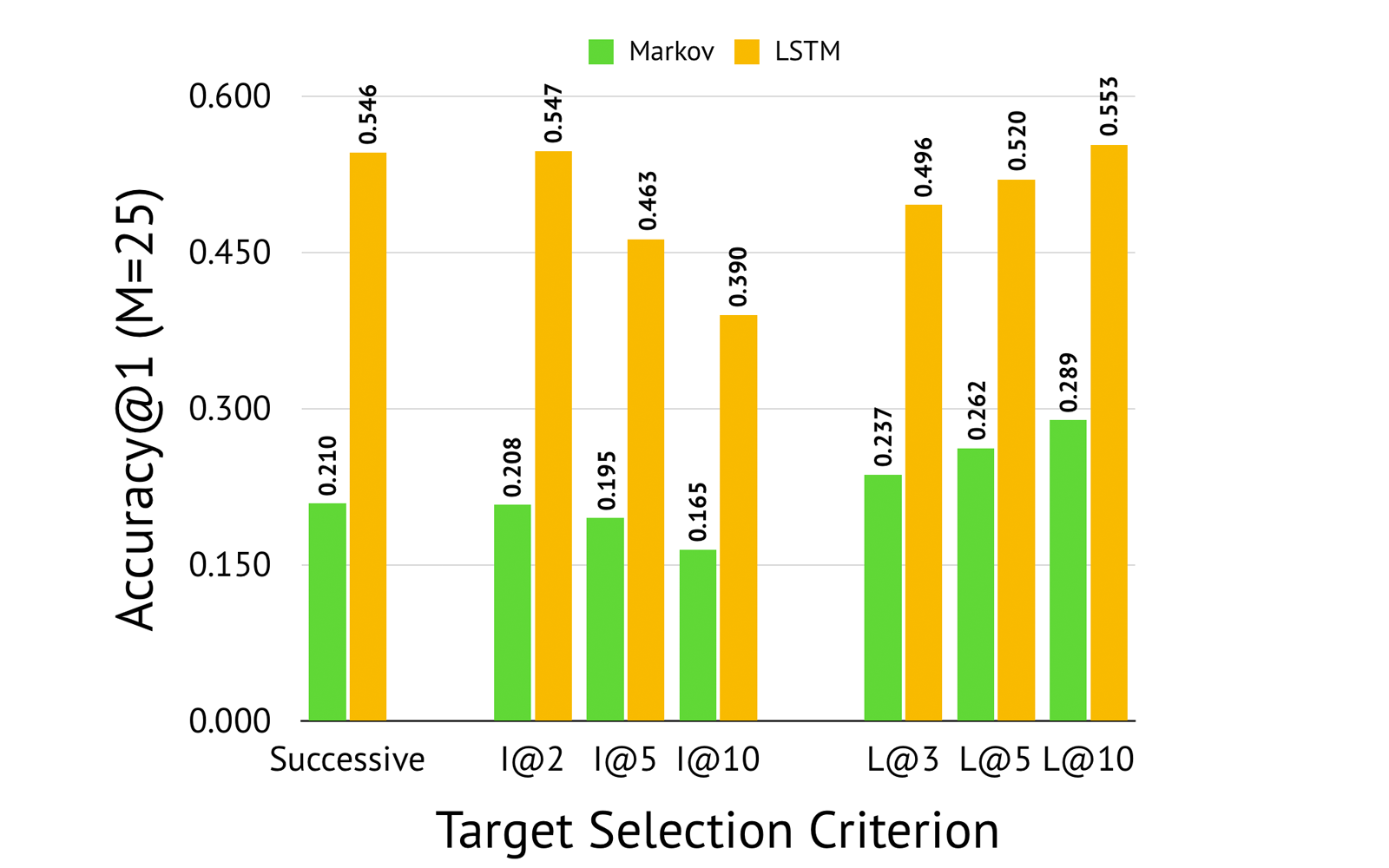}
		\caption[7.5pt]{Prediction accuracy under all target selection criteria. We show only the results for $M=25$. To save the space, we use ``I'' to denote Important and ``L'' to denote Longest.}
		\label{fig:result-target}
	\end{center}
\end{figure}

The prediction accuracy of the two models under $M=25$ is shown in Figure~\ref{fig:result-target}. For clarity, we show the results under only this location granularity. According to the figure, the performance of the Markov model and the LSTM model present the same trend, so for simplicity we discuss only the results of the LSTM model (yellow bars). We first concentrate on the performance of Important@$K$. The accuracy is lower with a larger threshold $K$. When the threshold of the important location is two minutes, the prediction accuracy is 0.547. When this threshold is 10 minutes, the accuracy degrades to 0.390. The results convey to us that if the standard for a important location is higher, it is harder to make an accurate prediction. 

The above results lead us to this conclusion: if the staying time of the target location is longer, the target location is harder to be accurately predicted. However, this is not consistent with the result of Longest@$K$. It is not hard to understand that the longest staying time of the first $K$ locations in $T_{w_f}^u$ will always increase with the increase of $K$. In our data, the average staying time of the target location when $M=25$ under Longest@3, Longest@5, and Longest@10 are 22.3 minutes, 36.3 minutes, and 53.1 minutes, respectively. From Figure~\ref{fig:result-target}, we can see that the prediction accuracy of Longest@$K$ is increasing with the increment of $K$. Hence, under such a circumstance, the target location with a longer staying time is easier to be predicted. From this point view, we can find that \textbf{the predictability is not simply correlated to the length of the staying time of the target location}. It is more affected by the form of the target selection criterion. 

Based on the preceding results, we claim that \textbf{Longest@$K$ should be the better criterion compared to Important@$K$}. We can choose a large $K$, by which we can better predict the real destination of the user. 

\section{Behavioral Feature Analysis}
\label{sec:feature}

The above sections have shown the impact of location granularity and location salience by stayting time. Finally, in this section, we study the prediction power of multiple types of behavioral features. We first describe how to extract behavioral features from the raw dataset, then introduce how to merge usage features into the LSTM model, and finally analyze the experiment results.

\subsection{Involved Features}

As stated in Section~\ref{sec:usage-data}, in this study we use the app usage data, the location sensor data, and the broadcasts to illustrate the user's behavior. In addition, we use the time information of an query to describe the temporal context of the user's usage. For each query, we calculate several statistics about the four groups of features within $w_h$ (the time window that corresponds to $T_{w_h}^u$). The brief introductions are listed as follows:

\begin{enumerate}
    \item \textit{App Usage.} We count which apps have been used at least once (either in foreground or background) within $w_h$. Since there are 655 apps in our raw dataset, we use a 655-dimensional binary vector to organize each data entry.
    \item \textit{Location Sensor Readings.} We calculate the average value of each sensor's reading within $w_h$. There are 238 sensors in this group of data, so this part of features is represented by a 238-dimensional real-value vector.
    \item \textit{Broadcasts.} We use a 82-dimensional vector to represent the broadcast information. Each dimension of the vector records the number of times that a broadcast occurred within $w_h$.
    \item \textit{Time.} For the temporal context, we roughly use two values: the beginning time and the ending time of $w_h$. For each of the value, we calculate which hour of the day and which day of the week it belongs to.
\end{enumerate}

\subsection{Performance of Each Single Group of Features}

\begin{figure}
	\centering
	\begin{center}
		\includegraphics[width=0.41\textwidth]{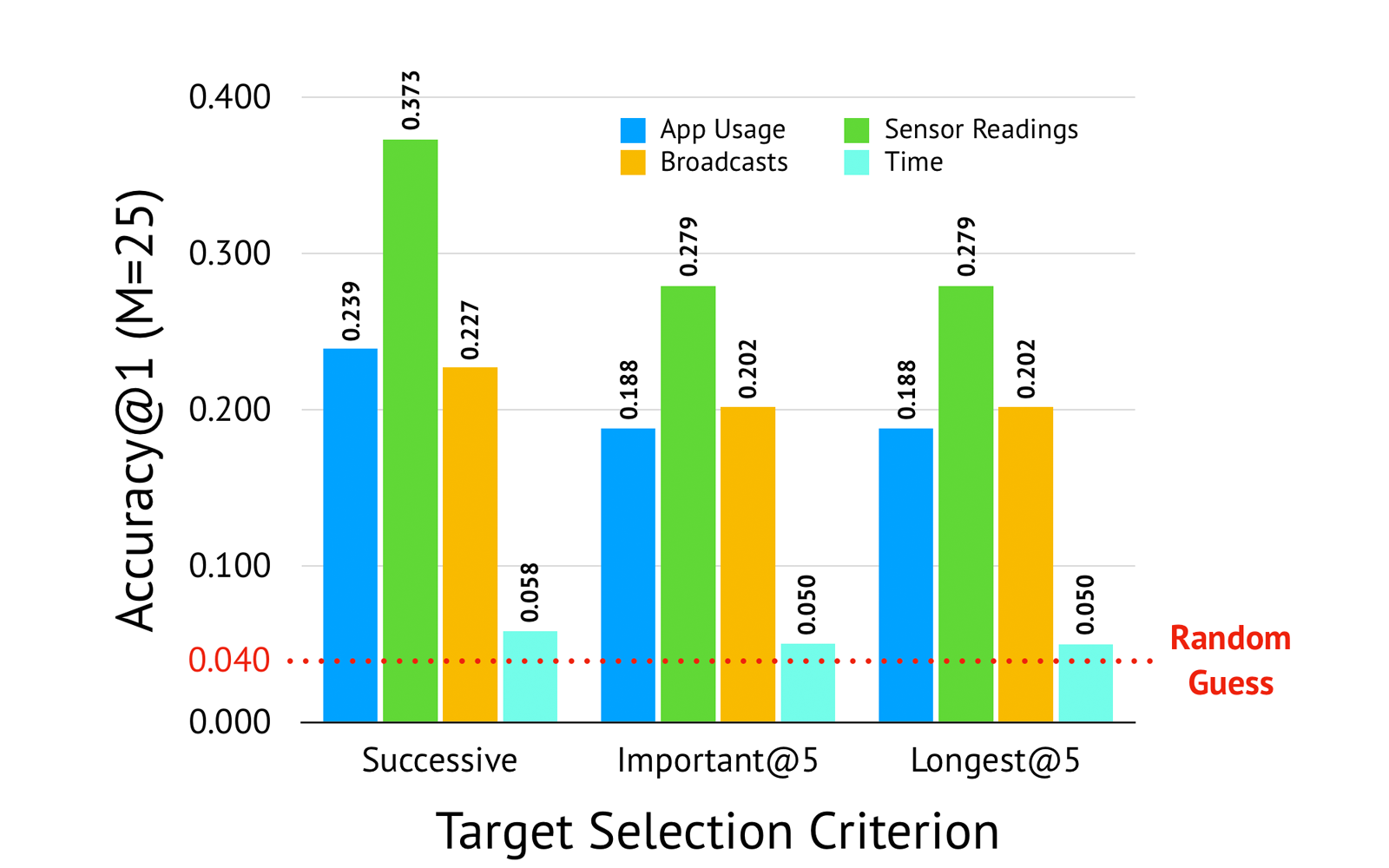}
		\caption[7.5pt]{Prediction accuracy of each single group of feature under Successive, Important@5, and Longest@5 under $M$=25. The red dotted line represents the accuracy of random guess.}
		\label{fig:result-single-feature}
	\end{center}
\end{figure}

\begin{figure}
    \centering
    \begin{center}
        \subfigure[The end-to-end neural network\label{fig:multimodal-1}]
        {\includegraphics[width=0.3\textwidth]{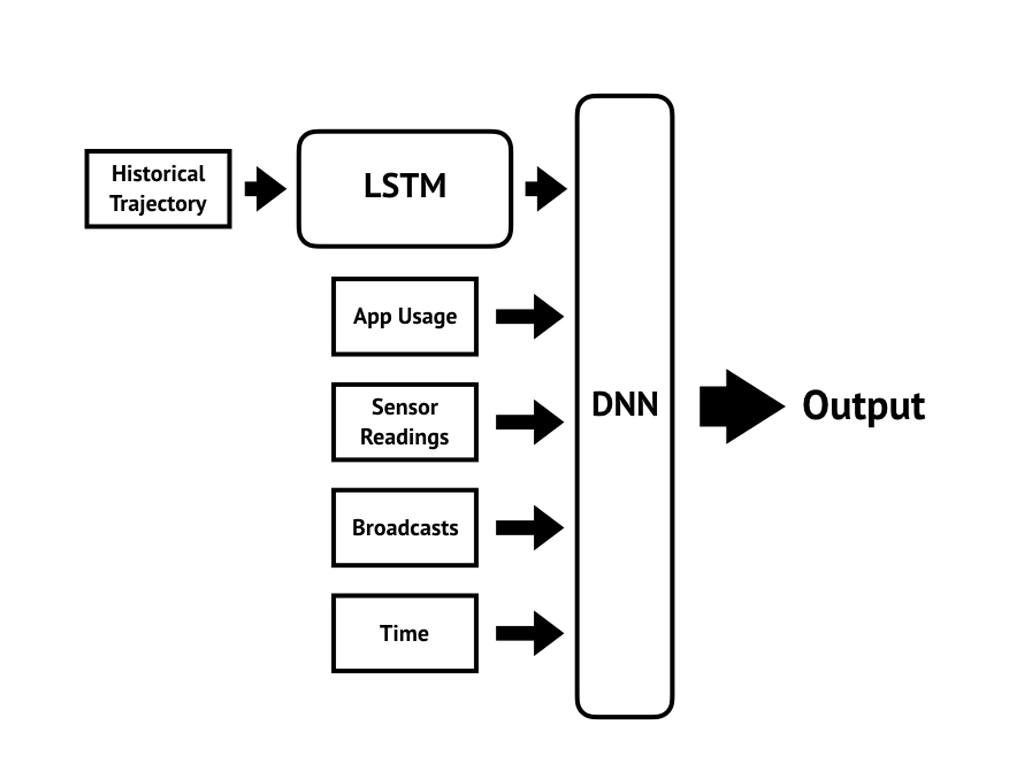}}
        \subfigure[The two-stage neural network\label{fig:multimodal-2}]
        {\includegraphics[width=0.3\textwidth]{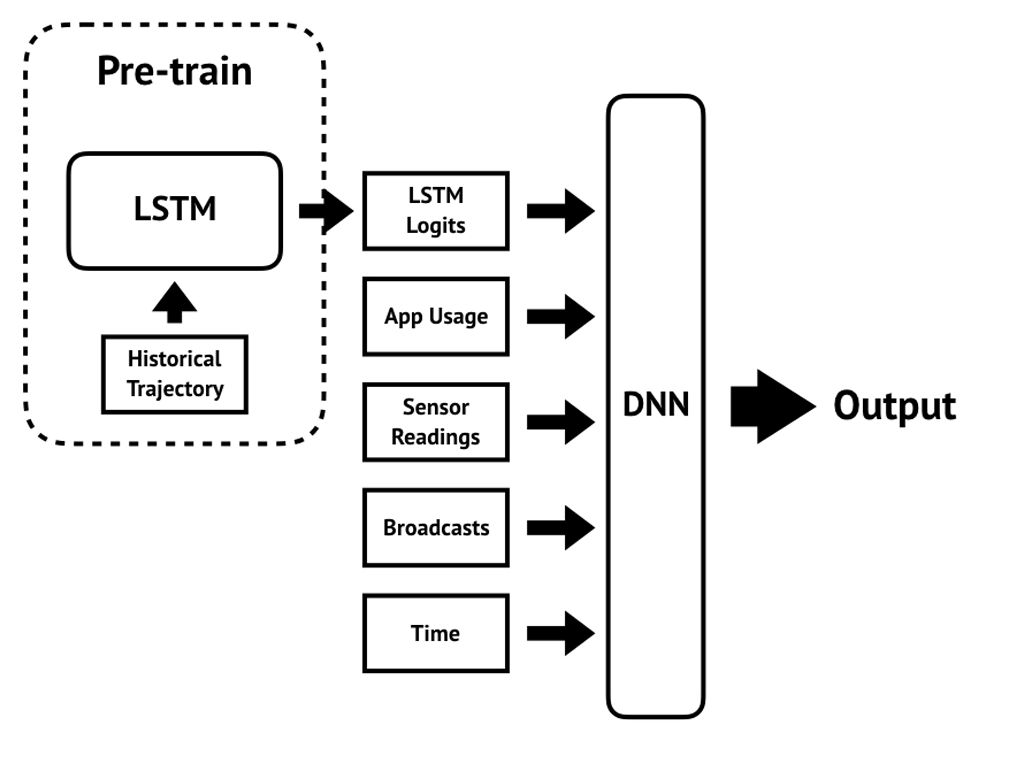}}
        \subfigure[Pre-trained neural network + Random Forest\label{fig:multimodal-3}]
        {\includegraphics[width=0.3\textwidth]{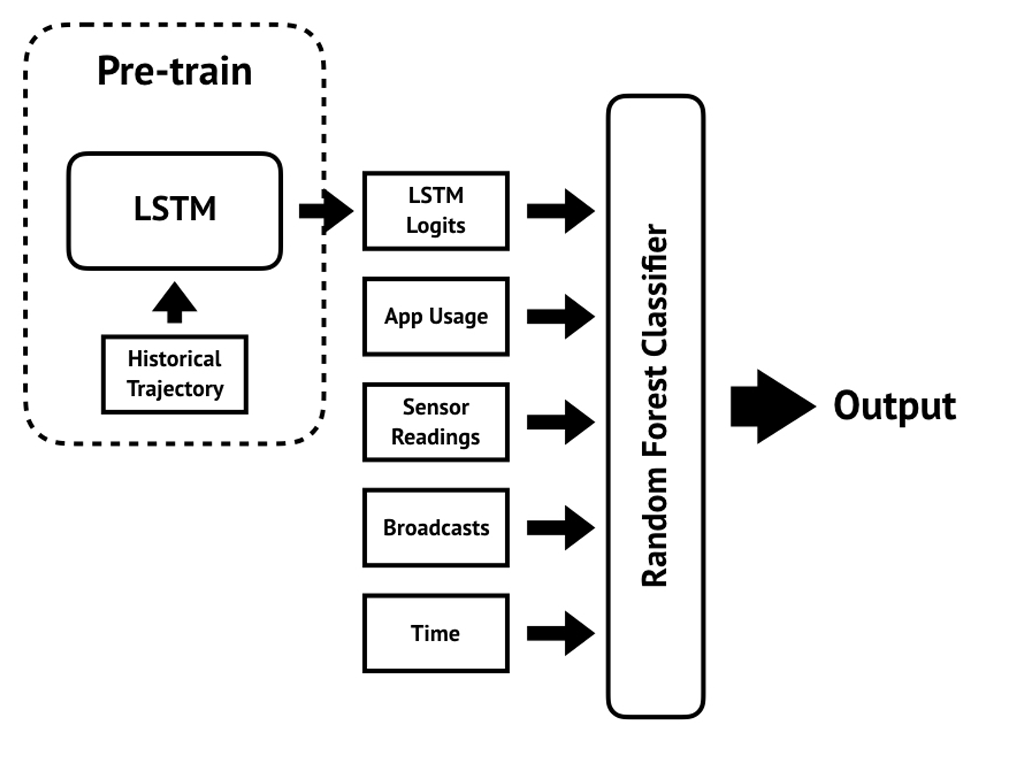}}
        \caption[7.5pt]{The three ways that we tried to integrate usage features into the LSTM model.}\label{fig:multimodal}
    \end{center}
\end{figure}

To examine the significance of features, we first attempt to predict the user's movement purely based on each single group of feature. During this step, we adopt a Random Forest Classifier~\cite{ho1995random} as the prediction model. To make the expression more clear, we denote the combination of a location granularity $G$ and a target selection criterion $C$ as a \textbf{scenario}. The results of three selected scenarios (Successive, Important@$5$, and Longest@$5$) under $M=25$ are shown in Figure~\ref{fig:result-single-feature}. 

The relative relationships among all groups of features are similar under all scenarios. Sensor readings could produce the best results, while app usage and broadcasts could get similar performance. All of these three groups of features could significantly outperform the random guess. From these results, we can learn that location sensors' readings are highly correlated to users' movement in the future. This is consistent with our common knowledge. A simple example is, if we know that the user is moving at high speed by observing the accelerometer, then we can guess that the user is driving to a location that is far away. As for app usage and broadcasts, we claim that they can provide useful information for the mobility prediction because they can somehow reflect users' usage behavior in the past, and the usage behavior is relevant to the user's movement in the future. Different from the location sensors, they do not have that close correlations with the user's movement, so they can not contribute as much as location sensors.

Although time is the weakest feature group, it still can outperform the random guessing a little bit. This indicates that knowing the current time can help predicting the user's next move, but the contribution is not that significant. In our common knowledge, time should be a very important factor that affects users' movement, since users usually have a fixed daily schedule. However, our results reveal that the users' movement at a fixed time is still quite uncertain.

\subsection{The Multi-Modal Learning}

\begin{figure}
	\centering
	\begin{center}
		\includegraphics[width=0.41\textwidth]{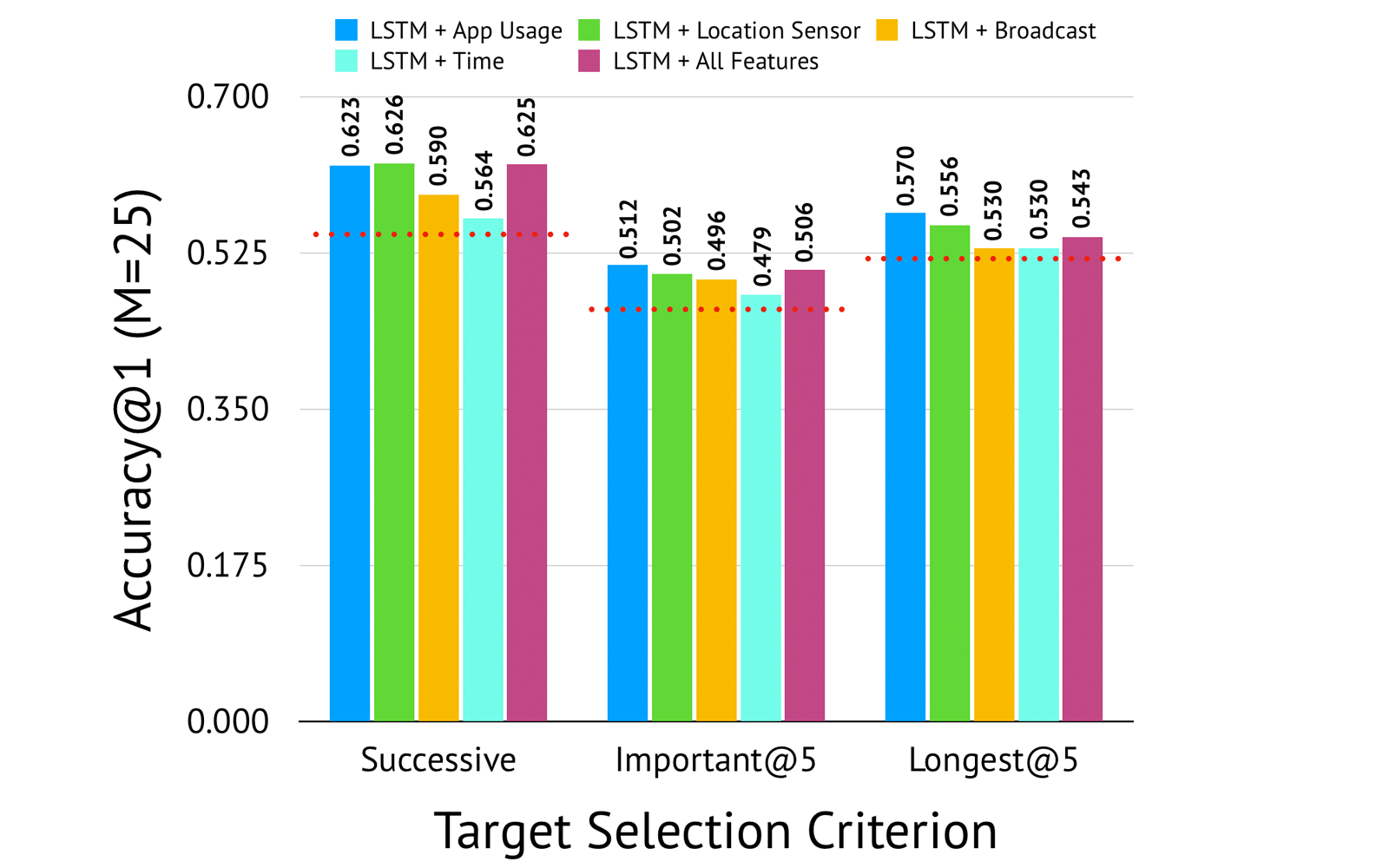}
		\caption[7.5pt]{Prediction accuracy of each combination of features under Successive, Important@5, and Longest@5 under $M$=25. The black lines represent the accuracy of the trajectory-based LSTM model.}
		\label{fig:result-multimodal}
	\end{center}
\end{figure}

Although the individual behavioral features are demonstrated to be rather effective, none of them can generate a better performance compared to the trajectory-based LSTM model. Indeed, the user's movement should be the main feature for the mobility prediction, while other features may provide only auxiliary information. Therefore, we need to find a way to synthesize  behavioral features into the LSTM model in order to achieve a better result.

The most straightforward way is using the Deep Neural Network (DNN) model to combine all these features. As demonstrated in Figure~\ref{fig:multimodal-1}, the historical trajectory is first encoded by the LSTM model before it is fed into the DNN. However, this model has the similar performance compared to  the trajectory-based LSTM, indicating that it cannot capture the information in the usage features. The reason could reside that the DNN is dominated by the strongest signal (historical trajectory), and neglects the other weak features. We then tried to pre-train a LSTM based on the historical trajectory in advance, and use the logit of that LSTM as a new group of feature (Figure~\ref{fig:multimodal-2}). Unfortunately, this model still does not produce any  improvement. According to these phenomenon, we argue that the problem is due to the capability that DNN cannot make good use of the information in the weak signals. So eventually, we replace the DNN with the Random Forest Classifier (Figure~\ref{fig:multimodal-3}). This modal finally generates results that are better than the LSTM model. 

\subsubsection{Macro Results}

\begin{figure}
	\centering
	\begin{center}
		\includegraphics[width=0.48\textwidth]{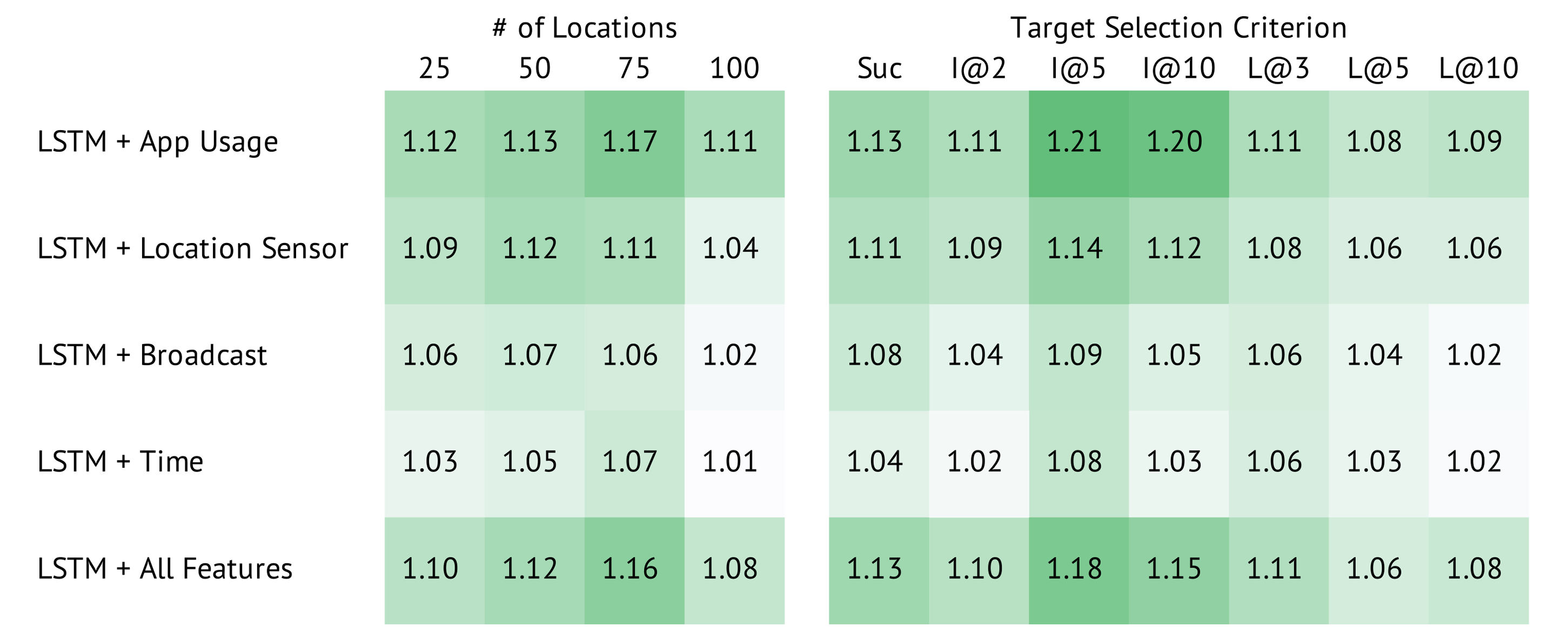}
		\caption[7.5pt]{The average relative performance of usage features under different location granularities and different target selection criteria.}
		\label{fig:heatmap}
	\end{center}
\end{figure}

We first combine each group of usage features with the LSTM logits to train the classifier, then use all the four groups together with the LSTM logits to train a comprehensive model. The results of the mentioned three scenarios (Successive, Important@$5$, and Longest@$5$) are shown in Figure~\ref{fig:result-multimodal}. First of all, we can see that each group of usage features can make contribution to the mobility prediction. The prediction accuracy can always be enhanced by involving a group of usage features. When comparing the significance from the four groups of features, it is interesting to notice that the most significant feature is app usage. Although sensor reading is the most effective group when used alone, it is not the best one anymore when jointly used with the LSTM model. This is because the information contained by location sensors and the historical trajectory are highly overlapped. The users' movement status described by location sensors are mostly contained in the historical trajectories as well, so integrating location sensors into the historical trajectories cannot provide additional information. Different from this, although app usage contains less information than location sensors, the knowledge that it contains cannot be contained by the historical trajectories. In other words, although it provides less information, it still has uniqueness that cannot be obtained from the historical trajectories. Therefore, it is easy to understand that synthesizing app usage into the LSTM model can make a better performance.

If we look at the performance when using all groups of features, we can see that \textbf{the prediction accuracy is not significantly better than integrating only app usage}. According to the above ideas, location sensors, broadcasts, and time almost cannot offer anything new. As a conclusion, we can learn from the results that app usage is the most important group of usage feature.

\subsubsection{Micro Results}

In the preceding analysis, we have shown that user usage features can contribute to the mobility prediction by analyzing three specific scenarios, which motivates us to explore the significance of these features under more scenarios. Since the performance of the pure LSTM model under different scenarios can vary a lot, to make a fair comparison, we define \textit{relative performance} as the ratio between the accuracy of the synthesized model with the accuracy of the pure LSTM model, and use relative performance to indicate the significance of the usage features under different scenarios.

Since it is hard to visualize a three-dimensional result tensor, we illustrate the average relative performance under each location granularity and each prediction target in Figure~\ref{fig:heatmap}. Intuitively, the darker the block's color is, the larger the value it holds. When calculating these average values, we do not include results under $M=5$ or $M=10$ since the numbers of valid data are too small (see Table~\ref{tab:testing-size}). 

From Figure~\ref{fig:heatmap}, we can obtain the following conclusions. First, the relative relationship among different features remains basically unchanged. App usage is always the most effective feature, and the location sensor is close behind. Broadcast and time can make only a little contribution. In other words, there is no feature that is particularly effective in a particular scenario. Second, if we focus on the left part of the figure, we can see that the effect of the app usage gets better as the number of locations increases, but drops if the number of locations is 100. Therefore, we can say that the app usage is most significant when the number of locations is moderate. In contrast, the effect of other features under different number of locations is not much different. Finally, from the right part of Figure~\ref{fig:heatmap}, we can learn that the usage features can make more contributions under Important@$K$ than Longest@$K$. This means the next important location is more relevant to the users' usage.

\section{Summary of Findings and Implications}
\label{sec:implication}

Finally, in this section, we summarize the key findings presented by the above analyses, and propose related implications for future research and development. 

First of all, we can see that \textbf{location granularity and salience do have significant impacts on mobility prediction}. Such impacts have never been revealed by existing literature, but have to draw more attention in practice. As for location granularity, when it is moderate, a ``location'' is more likely to be consistent with our common knowledge. In addition, the prediction accuracy reaches the peak when the location granularity is moderate, too. This conveys to us that \textit{when designing a mobility prediction model, the developer should carefully consider which location granularities are proper according to their scenario}, in order to make the prediction more effective and more practical. It is less likely to generate a usable prediction model at a improper location granularity.

As for location salience, we could see that the prediction performance varies quite significantly under different target selection criteria. When taking into account location salience, the performance is degraded compared to predict the exact next location. From this point of view, although existing models could predict the exact next location at a high accuracy, they may not have the ability to predict the next salient location well. Moreover, through our experiments, we found that the prediction performance is not simply correlated to the location salience. It is more affected by the form of the criterion. As a result, \textit{Longest@$K$ is the best selection criterion}, because we have more chance to locate the real destination of the user under this criterion, and we can also get better prediction performance at the same time.

Eventually, we have demonstated that behavioral features are quite indicative for the mobility prediction task. When being used individually, we found that location sensors are the most predictive behavioral features, because they are more relevant to the user's movement. However, if being synthesized with historical trajectories, app usage becomes the most helpful part of features, because it can offer additional information that the historical trajectories do not contain. Therefore, if a developer (e.g., the Location-Based Service provider) tries to build a mobility prediction model that relies on only behavioral data, \textit{she should consider location-related features (e.g., location sensors) as her first choice}. On the contrary, if the developer wants to utilize both behavioral features and historical trajectories, \textit{she should pay more attention to features that are more close to the user's active usage (e.g. app usage).}
\section{Conclusion and Future Work}
\label{sec:conclusion}

In this paper, we have rethought the human mobility prediction from a new perspective. That is, how will the problem setup influence the prediction performance. We designed a general measurement pipeline that can be used to analysis the issues, and conducted a comprehensive case study based on the Sherlock dataset to show the effectiveness of our pipeline. During the case study, we found many interesting results, and finally brought up several implications based on the results. Both the results and the implications are quite helpful for the future research in this area. In the future, we plan to explore more practical scenarios based on the findings of this article, and improve the effect of the mobility prediction models accordingly.

\bibliographystyle{unsrt}
\bibliography{references}

\end{document}